\documentstyle[seceq,preprint]{ptptex}

\newcommand{\second}{I\hspace{-1mm}I} 

\preprintnumber{TOYAMA-92
}
\title{
Massless Limits of Massive Tensor Fields ~\second \\
{\it --- Infrared regularization of Fierz-Pauli model ---}
}

\author{
Shinji {\sc Hamamoto}
\footnote{E-mail address: hamamoto@sci.toyama-u.ac.jp}
}

\inst{
Department of Physics, Toyama University, Toyama 930
}

\recdate{
\today
}

\abst{%
Izawa's gauge-fixing 
procedure based on BRS symmetry is applied twice to 
the massive tensor field theory of Fierz-Pauli type. 
It is shown the second application can remove massless 
singularities  which remain after the first application.
Massless limit of the theory is discussed. 
}

\begin{document}

\maketitle

\section{Introduction}
In a previous paper \cite{rf:1} (referred to as I), 
we investigated how 
Izawa's gauge-fixing procedure \cite{rf:2} based on BRS symmetry 
works 
for infrared regularization of massive tensor fields. 
We studied two models for a linearized massive tensor field, the 
pure-tensor (PT) type model by Fierz-Pauli and the 
additional-scalar-ghost (ASG) type one. 
It turned out that Izawa's procedure is effective for the ASG 
model, but not for the PT model. 
In the case of the ASG model, Izawa's procedure can regularize 
the original massless singularities of second order. 
On the other hand, the original singularities contained in the 
PT model are of fourth order, and Izawa's procedure can only 
reduce them to second order.

What we have learned from the above exercise is that when 
Izawa's procedure is applied once, massless singularities are 
reduced by second order. 
Now comes the question what happens when we apply Izawa's 
procedure once more to the PT model. 
This is the issue to be discussed in the present paper. 
We show that the second application of Izawa's procedure 
does regularize the remaining second-order massless singularities 
in the PT model. 

In \S 2, the results obtained in I concerning the first application 
of Izawa's procedure are summarized.
In \S 3, the second application of Izawa's procedure is 
performed to the PT model. 
It is shown that the second-order massless singularities which 
remain after the first application are in fact regularized. 
In \S 4, we discuss massless limit of the resulting theory. 
We see a graviton field is contained in the theory as a nonlocal 
combination of the basic fields. 
Summary and discussion are given in \S 5. 
In Appendix we show that applying Izawa's procedure to a 
massive vector field reproduces the Stueckelberg formalism.

\section{First application of Izawa's procedure}

A massive tensor field is described by the Lagrangian
\footnote{Notations used in this paper are the same as in I.}
\begin{equation}
L_{h} = 
\frac{1}{2}h^{\mu\nu}\Lambda_{\mu\nu ,\rho\sigma}h^{\rho\sigma} 
- \frac{m^{2}}{2}\left( h^{\mu\nu}h_{\mu\nu} - ah^{2}\right) ,
\label{eq:201}
\end{equation}
where $\Lambda_{\mu\nu ,\rho\sigma}$ is the operator defined by 
\begin{eqnarray}
\Lambda_{\mu\nu ,\rho\sigma} & = & 
\left(\eta_{\mu\rho}\eta_{\nu\sigma} 
- \eta_{\mu\nu}\eta_{\rho\sigma}\right)\Box 
\nonumber \\
& & \mbox{}- \left(\eta_{\mu\rho}\partial_{\nu}\partial_{\sigma} 
+ \eta_{\nu\sigma}\partial_{\mu}\partial_{\rho}\right)
+ \left(\eta_{\rho\sigma}\partial_{\mu}\partial_{\nu} 
+ \eta_{\mu\nu}\partial_{\rho}\partial_{\sigma}\right) ,
\end{eqnarray}
and $a$ is a real parameter taking the values 
\begin{equation}
a = \left\{
\begin{array}{ll}
\frac{1}{2} & {\rm for \ the \ ASG \ model} , \\
1           & {\rm for \ the \ PT \ model} .
\end{array}
\right.
\end{equation}
Two-point functions are calculated as 
\begin{eqnarray}
\langle h^{\mu\nu}h^{\rho\sigma}\rangle & = & 
\frac{1}{\Box - m^{2}}\left\{
\frac{1}{2}\left(
\eta^{\mu\rho}\eta^{\nu\sigma} 
+ \eta^{\mu\sigma}\eta^{\nu\rho} 
- \eta^{\mu\nu}\eta^{\rho\sigma}\right) \right. \nonumber \\
& & \makebox[10mm]{}\left.\mbox{}
- \frac{1}{2m^{2}}\left(
\eta^{\mu\rho}\partial^{\nu}\partial^{\sigma}
+ \eta^{\mu\sigma}\partial^{\nu}\partial^{\rho}
+ \eta^{\nu\rho}\partial^{\mu}\partial^{\sigma}
+ \eta^{\nu\sigma}\partial^{\mu}\partial^{\rho}\right)\right\}\delta
\nonumber \\
& & \label{eq:204}
\end{eqnarray}
for the ASG model, and
\begin{eqnarray}
\langle h^{\mu\nu}h^{\rho\sigma}\rangle & = & 
\frac{1}{\Box - m^{2}}\left\{
\frac{1}{2}\left(
\eta^{\mu\rho}\eta^{\nu\sigma} 
+ \eta^{\mu\sigma}\eta^{\nu\rho} 
- \eta^{\mu\nu}\eta^{\rho\sigma}\right) \right. \nonumber \\
& & \makebox[15mm]{}
- \frac{1}{2m^{2}}\left(
\eta^{\mu\rho}\partial^{\nu}\partial^{\sigma}
+ \eta^{\mu\sigma}\partial^{\nu}\partial^{\rho}
+ \eta^{\nu\rho}\partial^{\mu}\partial^{\sigma}
+ \eta^{\nu\sigma}\partial^{\mu}\partial^{\rho}\right) \nonumber \\
& & \makebox[14mm]{}\left.\mbox{}
+ \frac{2}{3}\left(
\frac{1}{2}\eta^{\mu\nu} 
+ \frac{\partial^{\mu}\partial^{\nu}}{m^{2}}\right)\left(
\frac{1}{2}\eta^{\rho\sigma} 
+ \frac{\partial^{\rho}\partial^{\sigma}}{m^{2}}\right)
\right\}\delta
\label{eq:205}
\end{eqnarray}
for the PT model.
We see that the ASG model has second-order massless 
singularities, while the PT model fourth-order.

Applying Izawa's gauge-fixing procedure based on BRS symmetry, 
we obtain the following Lagrangian:
\begin{eqnarray}
L_{\rm T} & = & 
\frac{1}{2}h^{\mu\nu}\Lambda_{\mu\nu ,\rho\sigma}h^{\rho\sigma} 
\nonumber \\
& & \mbox{}
- \frac{m^{2}}{2}\left[
\left(h_{\mu\nu} - \frac{1}{m}\left(
\partial_{\mu}\theta_{\nu} + \partial_{\nu}\theta_{\mu}
\right)\right)^{2} 
- a\left( h - \frac{2}{m}\partial^{\mu}\theta_{\mu}\right)^{2}
\right] \nonumber \\
& & \mbox{}
+\, b^{\mu}\left(
\partial^{\nu}h_{\mu\nu} - \frac{1}{2}\partial_{\mu}h 
+ \frac{\alpha}{2}b_{\mu}\right) 
+ i\bar{c}^{\mu}\Box c_{\mu} ,
\label{eq:206}
\end{eqnarray}
where an auxiliary vector field $\theta_{\mu}$, 
a Nakanishi-Lautrup (NL) field $b_{\mu}$, 
a pair of Faddeev-Popov (FP) ghosts $(c_{\mu}, \bar{c}_{\mu})$, and 
a gauge parameter $\alpha$ 
have been introduced. 
This Lagrangian is invariant under the following BRS transformation:
\begin{equation}
\begin{array}{lclcl}
\delta h_{\mu\nu} = \partial_{\mu}c_{\nu} + \partial_{\nu}c_{\mu} , 
& & 
\delta\theta_{\mu} = mc_{\mu} , 
& & 
\delta\bar{c}_{\mu} = ib_{\mu} . 
\end{array}
\end{equation}

Putting $a=\frac{1}{2}$ in Eq.(\ref{eq:205}), we have for the 
ASG model
\begin{eqnarray}
L_{{\rm T},a=\frac{1}{2}} & = & 
\frac{1}{2}h^{\mu\nu}\Lambda_{\mu\nu ,\rho\sigma}h^{\rho\sigma} 
\nonumber \\
& & \mbox{}
- \frac{m^{2}}{2}\left( h^{\mu\nu}h_{\mu\nu} 
- \frac{1}{2}h^{2}\right) 
- 2m\theta^{\mu}\left(
\partial^{\nu}h_{\mu\nu} - \frac{1}{2}\partial_{\mu}h\right) 
- \partial_{\mu}\theta_{\nu}\partial^{\mu}\theta^{\nu} \nonumber \\
& & \mbox{}
+ \, b^{\mu}\left(
\partial^{\nu}h_{\mu\nu} - \frac{1}{2}\partial_{\mu}h 
+ \frac{\alpha}{2}b_{\mu}\right) 
+ i\bar{c}^{\mu}\Box c_{\mu} . \label{eq:208}
\end{eqnarray}
This gives the following two-point functions:
\begin{eqnarray}
\langle h^{\mu\nu}h^{\rho\sigma}\rangle & = & 
\frac{1}{\Box - m^{2}}\left\{
\frac{1}{2}\left(
\eta^{\mu\rho}\eta^{\nu\sigma} 
+ \eta^{\mu\sigma}\eta^{\nu\rho} 
- \eta^{\mu\nu}\eta^{\rho\sigma}\right)
\begin{array}{ll}
 & \\
 & 
\end{array}
\right. \nonumber \\
& & \makebox[-15mm]{}\left.\mbox{}
- \frac{1}{2}\left[
(1-2\alpha )\frac{1}{\Box} 
+ 2\alpha\frac{m^{2}}{\Box^{2}}\right]\left(
\eta^{\mu\rho}\partial^{\nu}\partial^{\sigma}
+ \eta^{\mu\sigma}\partial^{\nu}\partial^{\rho}
+ \eta^{\nu\rho}\partial^{\mu}\partial^{\sigma}
+ \eta^{\nu\sigma}\partial^{\mu}\partial^{\rho}\right)
\right\}\delta , 
\nonumber \\ 
& & \\
\langle h^{\mu\nu}b^{\rho}\rangle & = & 
\frac{1}{\Box}\left( 
\eta^{\mu\rho}\partial^{\nu} 
+ \eta^{\nu\rho}\partial^{\mu}\right)\delta , \\
\langle h^{\mu\nu}\theta^{\rho}\rangle & = & 
\mbox{}- \alpha m\frac{1}{\Box^{2}}\left(
\eta^{\mu\rho}\partial^{\nu} + \eta^{\nu\rho}\partial^{\mu}\right) 
\delta , \\
\langle b^{\mu}b^{\rho}\rangle & = & 0 , \\
\langle b^{\mu}\theta^{\rho}\rangle & = & 
m\frac{1}{\Box}\eta^{\mu\rho}\delta , \\
\langle\theta^{\mu}\theta^{\rho}\rangle & = & 
\frac{1}{2}\frac{1}{\Box}\left(
1 - 2\alpha\frac{m^{2}}{\Box}\right)\eta^{\mu\rho}\delta ,
\label{eq:214}
\end{eqnarray}
except for the trivial one for 
$\langle c^{\mu}\bar{c}^{\nu}\rangle$.
\footnote{
In Abelian cases which we are dealing with, FP ghosts decouple 
from all other fields to give trivial two-point functions. 
We omit to write down their explicit forms throughout the paper.
}
We see that the massless singularities in (\ref{eq:204}) have 
been regularized by this procedure.
\label{note1}
\footnote{
Exactly speaking, the quantity $\Box^{-2}$ is well-defined 
only when accompanied by derivatives. 
Therefore, the expression (\ref{eq:214}) is meaningful only in 
a formal sense.
If the field $\theta_{\mu}$ has some non-derivative couplings, 
$\alpha$ is to be set 0. 
On the other hand, if it has derivative couplings only, then any 
value of $\alpha$ is allowed.
}

When $a=1$, the Lagrangian (\ref{eq:206}) gives
\begin{eqnarray}
L_{{\rm T},a=1} & = & 
\frac{1}{2}h^{\mu\nu}\Lambda_{\mu\nu ,\rho\sigma}h^{\rho\sigma} 
\nonumber \\
& & \mbox{}
- \frac{m^{2}}{2}\left( h^{\mu\nu}h_{\mu\nu} - h^{2}\right) 
- 2m\theta^{\mu}\left(\partial^{\nu}h_{\mu\nu} 
- \partial_{\mu}h\right) 
- \frac{1}{2}\left(
\partial_{\mu}\theta_{\nu} - \partial_{\nu}\theta_{\mu}\right)^{2}
\nonumber \\
& & \mbox{}
+\, b^{\mu}\left(
\partial^{\nu}h_{\mu\nu} - \frac{1}{2}\partial_{\mu}h 
+ \frac{\alpha}{2}b_{\mu}\right) 
+ i\bar{c}^{\mu}\Box c_{\mu} . \label{eq:215}
\end{eqnarray}
Thus two-point functions for the PT model are
\begin{eqnarray}
\langle h^{\mu\nu}h^{\rho\sigma}\rangle & = & 
\frac{1}{\Box - m^{2}}\left\{
\frac{1}{2}\left(
\eta^{\mu\rho}\eta^{\nu\sigma} 
+ \eta^{\mu\sigma}\eta^{\nu\rho} 
- \eta^{\mu\nu}\eta^{\rho\sigma}\right) \right. \nonumber \\
& & \makebox[-1cm]{}
- \frac{1}{2}\left[
(1-2\alpha )\frac{1}{\Box} 
+ 2\alpha\frac{m^{2}}{\Box^{2}}\right]\left(
\eta^{\mu\rho}\partial^{\nu}\partial^{\sigma}
+ \eta^{\mu\sigma}\partial^{\nu}\partial^{\rho}
+ \eta^{\nu\rho}\partial^{\mu}\partial^{\sigma}
+ \eta^{\nu\sigma}\partial^{\mu}\partial^{\rho}\right) \nonumber \\
& & \makebox[-1mm]{}\left.\mbox{}
+ \frac{2}{3}\left(
\frac{1}{2}\eta^{\mu\nu} 
+ \frac{\partial^{\mu}\partial^{\nu}}{\Box}\right)\left(
\frac{1}{2}\eta^{\rho\sigma} 
+ \frac{\partial^{\rho}\partial^{\sigma}}{\Box}\right)
\right\} \delta ,\\ 
\langle h^{\mu\nu}b^{\rho}\rangle & = & 
\frac{1}{\Box}\left( 
\eta^{\mu\rho}\partial^{\nu} 
+ \eta^{\nu\rho}\partial^{\mu}\right)\delta , \\
\langle h^{\mu\nu}\theta^{\rho}\rangle & = & 
\left\{
\frac{1}{6m}\frac{1}{\Box}\eta^{\mu\nu}\partial^{\rho} 
- \alpha m\frac{1}{\Box^{2}}\left(
\eta^{\mu\rho}\partial^{\nu} + \eta^{\nu\rho}\partial^{\mu}\right) 
+ \frac{1}{3m}\frac{1}{\Box^{2}}
\partial^{\mu}\partial^{\nu}\partial^{\rho}
\right\}\delta , 
\label{eq:218} \\
\langle b^{\mu}b^{\rho}\rangle & = & 0 , \\
\langle b^{\mu}\theta^{\rho}\rangle & = & 
m\frac{1}{\Box}\eta^{\mu\rho}\delta , \\
\langle\theta^{\mu}\theta^{\rho}\rangle & = & 
\left\{
\frac{1}{2}\frac{1}{\Box}\left(
1 - 2\alpha\frac{m^{2}}{\Box}\right)\eta^{\mu\rho} 
- \frac{1}{6m^{2}}\frac{1}{\Box}\left(
1 - \frac{m^{2}}{\Box}\right)
\partial^{\mu}\partial^{\rho}\right\}\delta .
\label{eq:221}
\end{eqnarray}
We see the fourth-order massless singularities found 
in (\ref{eq:205}) for 
the original PT model have been driven away. 
However, new at-most-second-order singularities have 
appeared in the $\theta$-sector (\ref{eq:218}) and (\ref{eq:221}).
Can these singularities be driven away by applying Izawa's 
procedure once more? 
This is the issue to address in the next section.

\section{Second application of Izawa's procedure}

The starting point is the Lagrangian (\ref{eq:215}). 
The kinetic term of the vector field $\theta_{\mu}$ shows this 
field is a kind of gauge field. 
It is expected from this fact that the second application of 
Izawa's procedure works. 
This is in fact the case as seen below.

We introduce a new set of variables $(\theta_{\mu}', \varphi)$ 
to perform a field transformation 
$\theta \rightarrow (\theta_{\mu}', \varphi)$ such that
\begin{eqnarray}
\theta_{\mu} & = & 
\theta_{\mu}' - \frac{1}{m}\partial_{\mu}\varphi , \\
\partial^{\mu}\theta_{\mu}' & = & 0 .
\end{eqnarray}
The new variables $(\theta_{\mu}', \varphi)$ are first assumed to 
be independent of the old one $\theta_{\mu}$.
Then the Lagrangian (\ref{eq:215}), which does not depend on the new 
variables, is invariant under the BRS transformation
\begin{equation}
\left\{
\begin{array}{lcl}
\delta'\theta_{\mu}' = c_{\mu}' , & & 
\delta' \bar{c}_{\mu}' = ib_{\mu}' , 
\\
\delta' \varphi = mc ,        & & 
\delta' \bar{c} = ib ,
\end{array}
\right.
\end{equation}
where the new FP ghosts 
$(c_{\mu}', c)$ and $(\bar{c}_{\mu}',\bar{c})$ 
as well as the new NL fields $(b_{\mu}', b)$ have been introduced. 
To relate the old and new sets of variables, 
we supplement the Lagrangian (\ref{eq:215}) by adding the 
following BRS gauge-fixing term: 
\begin{eqnarray}
L_{\rm B}' & = & 
\mbox{}-i\delta'\left[ 
\bar{c}^{\mu\prime}\left(\theta_{\mu} - \theta_{\mu}' + \frac{1}{m}
\partial_{\mu}\varphi\right) 
+ \bar{c}\left(\partial^{\mu}\theta_{\mu}' 
- \frac{m}{2}h 
+ \frac{\beta}{2}b\right)\right] 
\nonumber \\
& = & \,
b^{\mu\prime}\left(\theta_{\mu} - \theta_{\mu}' 
+ \frac{1}{m}\partial_{\mu}\varphi\right) 
+ b\left(\partial^{\mu}\theta_{\mu}' 
- \frac{m}{2}h 
+ \frac{\beta}{2}b\right) 
\nonumber \\
& & \makebox[15mm]{}
-i\left(\bar{c}^{\mu\prime} + \partial^{\mu}\bar{c}\right) 
\left(c_{\mu}' - \partial_{\mu}c\right) 
+ i\bar{c}\Box c 
\end{eqnarray}
with the second gauge parameter $\beta$. 
The path integral is given as
\begin{eqnarray}
Z & = & 
\int {\cal D}h_{\mu\nu}{\cal D}\theta_{\mu}
{\cal D}b_{\mu}{\cal D}c_{\mu}{\cal D}\bar{c}_{\mu}
{\cal D}\theta_{\mu}'{\cal D}\varphi
{\cal D}b_{\mu}'{\cal D}c_{\mu}'{\cal D}\bar{c}_{\mu}'
{\cal D}b{\cal D}c{\cal D}\bar{c} \nonumber \\
& & \makebox[15mm]{}
\times \exp i\int d^{4}x\left[ L_{{\rm T}, a=1}
+ L_{\rm B}'\right] .
\end{eqnarray}
Integrating over the variables 
$(b_{\mu}', \theta_{\mu}, c_{\mu}', \bar{c}_{\mu}')$ 
and overwriting $\theta_{\mu}$ on $\theta_{\mu}'$, we obtain 
\begin{equation}
Z = \int 
{\cal D}h_{\mu\nu}{\cal D}\theta_{\mu}{\cal D}\varphi
{\cal D}b_{\mu}{\cal D}b
{\cal D}c_{\mu}{\cal D}\bar{c}_{\mu}{\cal D}c{\cal D}\bar{c}
\exp i\int d^{4}x\, L_{{\rm T}, a=1}' ,
\end{equation}
where 
\begin{eqnarray}
L_{{\rm T}, a=1}' & = & 
\frac{1}{2}h^{\mu\nu}\Lambda_{\mu\nu ,\rho\sigma}
h^{\rho\sigma} 
\nonumber \\
& & \makebox[1cm]{}
- \frac{m^{2}}{2}\left[
\left(h_{\mu\nu} - \frac{1}{m}\left(
\partial_{\mu}\theta_{\nu} + \partial_{\nu}\theta_{\mu}\right)
+ \frac{2}{m^{2}}\partial_{\mu}\partial_{\nu}\varphi
\right)^{2} \right.
\nonumber \\
& & \makebox[25mm]{}\left.
- \left( h - \frac{2}{m}\partial^{\mu}\theta_{\mu}
+ \frac{2}{m^{2}}\Box\varphi
\right)^{2}
\right] \nonumber \\
& & \makebox[1cm]{}
+\, b^{\mu}\left(
\partial^{\nu}h_{\mu\nu} - \frac{1}{2}\partial_{\mu}h 
+ \frac{\alpha}{2}b_{\mu}\right) 
+ i\bar{c}^{\mu}\Box c_{\mu}
\nonumber \\
& & \makebox[1cm]{}
+\, b\left(
\partial^{\mu}\theta_{\mu} - \frac{m}{2}h 
+ \frac{\beta}{2}b\right)
+ i\bar{c}\Box c 
\nonumber \\
& = & 
\frac{1}{2}h^{\mu\nu}\Lambda_{\mu\nu ,\rho\sigma}h^{\rho\sigma} 
- \frac{m^{2}}{2}\left( h^{\mu\nu}h_{\mu\nu} - h^{2}\right) 
\nonumber \\
& & \makebox[1cm]{}
- 2\left( m\theta^{\mu} - \partial^{\mu}\varphi\right)
\left(\partial^{\nu}h_{\mu\nu} - \partial_{\mu}h\right) 
- \frac{1}{2}\left(
\partial_{\mu}\theta_{\nu} - \partial_{\nu}
\theta_{\mu}\right)^{2}
\nonumber \\
& & \makebox[1cm]{}
+\, b^{\mu}\left(
\partial^{\nu}h_{\mu\nu} - \frac{1}{2}\partial_{\mu}h 
+ \frac{\alpha}{2}b_{\mu}\right) 
+ i\bar{c}^{\mu}\Box c_{\mu}
\nonumber \\
& & \makebox[1cm]{}
+\, b\left(
\partial^{\mu}\theta_{\mu} - \frac{m}{2}h 
+ \frac{\beta}{2}b\right)
+ i\bar{c}\Box c .
\label{eq:307}
\end{eqnarray}
This Lagrangian is invariant under the following 
BRS transformation:
\begin{equation}
\left\{
\begin{array}{ll}
\delta h_{\mu\nu} = \partial_{\mu}c_{\nu} + 
\partial_{\nu}c_{\mu} , &
\delta\bar{c}_{\mu} = ib_{\mu} , \\
\delta\theta_{\mu} = mc_{\mu} + \partial_{\mu}c , &
\delta\bar{c} = ib , \\
\delta\varphi = mc . & 
\end{array}
\right.
\end{equation}
We note the Lagrangian $L_{{\rm T}, a=1}'$ has a 
smooth massless limit. 
This comes from the fact that the kinetic term of $\theta_{\mu}$
contained in 
the Lagrangian $L_{{\rm T}, a=1}$ (\ref{eq:215}) is 
the gauge-theoretic one 
$- \frac{1}{2}\left(
\partial_{\mu}\theta_{\nu} - \partial_{\nu}\theta_{\mu}\right)^{2}
$. 
On the contrary, the Lagrangian 
$L_{{\rm T}, a=\frac{1}{2}}$ (\ref{eq:208}) for the 
ASG model has the non-gauge-theoretic kinetic term 
$- \partial_{\mu}\theta_{\nu}\partial^{\mu}\theta^{\nu}$.
When performed the second application of Izawa's procedure, 
such a term yields at-most-second-order singular terms like 
$- \frac{1}{m^{2}}\left(\Box\varphi\right)^{2}$.

Two-point functions obtained from $L_{{\rm T}, a=1}'$ 
(\ref{eq:307}) are the following:
\begin{eqnarray}
\langle h^{\mu\nu}h^{\rho\sigma}\rangle & = & 
\frac{1}{\Box - m^{2}}\left\{
\frac{1}{2}\left(
\eta^{\mu\rho}\eta^{\nu\sigma} 
+ \eta^{\mu\sigma}\eta^{\nu\rho} 
- \eta^{\mu\nu}\eta^{\rho\sigma}\right) \right. \nonumber \\
& & \makebox[-1cm]{}
- \frac{1}{2}\left[
(1-2\alpha )\frac{1}{\Box} 
+ 2\alpha\frac{m^{2}}{\Box^{2}}\right]\left(
\eta^{\mu\rho}\partial^{\nu}\partial^{\sigma}
+ \eta^{\mu\sigma}\partial^{\nu}\partial^{\rho}
+ \eta^{\nu\rho}\partial^{\mu}\partial^{\sigma}
+ \eta^{\nu\sigma}\partial^{\mu}\partial^{\rho}\right) \nonumber \\
& & \makebox[-1mm]{}\left.\mbox{}
+ \frac{2}{3}\left(
\frac{1}{2}\eta^{\mu\nu} 
+ \frac{\partial^{\mu}\partial^{\nu}}{\Box}\right)\left(
\frac{1}{2}\eta^{\rho\sigma} 
+ \frac{\partial^{\rho}\partial^{\sigma}}{\Box}\right)
\right\} \delta ,
\label{eq:309} \\ 
\langle h^{\mu\nu}b^{\rho}\rangle & = & 
\frac{1}{\Box}\left( 
\eta^{\mu\rho}\partial^{\nu} 
+ \eta^{\nu\rho}\partial^{\mu}\right)\delta , 
\label{eq:310} \\
\langle h^{\mu\nu}b\rangle & = & 0 , 
\label{eq:311} \\
\langle h^{\mu\nu}\theta^{\rho}\rangle & = & \mbox{}
- \alpha m\frac{1}{\Box^{2}}\left(
\eta^{\mu\rho}\partial^{\nu} + \eta^{\nu\rho}\partial^{\mu}\right) 
\delta , 
\label{eq:312} \\
\langle h^{\mu\nu}\varphi\rangle & = & 
\frac{1}{3}\frac{1}{\Box}\left(
\frac{1}{2}\eta^{\mu\nu} 
+ \frac{\partial^{\mu}\partial^{\nu}}{\Box}\right)\delta , 
\label{eq:313} \\
\langle b^{\mu}b^{\rho}\rangle & = & 0 , 
\label{eq:314} \\
\langle b^{\mu}b\rangle & = & 0 ,
\label{eq:315} \\
\langle b^{\mu}\theta^{\rho}\rangle & = & 
m\frac{1}{\Box}\eta^{\mu\rho}\delta , 
\label{eq:316} \\
\langle b^{\mu}\varphi\rangle & = & 0 , 
\label{eq:317} \\
\langle bb\rangle & = & 0 , 
\label{eq:318} \\
\langle b\theta^{\mu}\rangle & = & \mbox{}
- \frac{1}{\Box}\partial^{\mu}\delta , 
\label{eq:319} \\
\langle b\varphi\rangle & = & 
m\frac{1}{\Box}\delta , 
\label{eq:320} \\
\langle\theta^{\mu}\theta^{\rho}\rangle & = & 
\left[
\frac{1}{2}\frac{1}{\Box}\left(
1 - 2\alpha\frac{m^{2}}{\Box}\right)\eta^{\mu\rho} 
- \frac{1}{2}\left( 1 - 2\beta\right)
\frac{\partial^{\mu}\partial^{\rho}}{\Box^{2}}\right]\delta ,
\label{eq:321} \\
\langle\theta^{\mu}\varphi\rangle & = & 
\frac{1}{2}\left( 1 - 2\beta\right)\frac{m}{\Box^{2}}
\partial^{\mu}\delta , 
\label{eq:322} \\
\langle\varphi\varphi\rangle & = & 
\left[\frac{1}{6}\frac{1}{\Box}\left(
1 - \frac{m^{2}}{\Box}\right)
+ \frac{1}{2}\left( 1 - 2\beta\right)
\frac{m^{2}}{\Box^{2}}\right]\delta .
\label{eq:323}
\end{eqnarray}
These expressions show that the massless singularities remaining in 
(\ref{eq:218}) and (\ref{eq:221}) have been regularized.
\footnote{
The note stated in the second footnote on p.~\pageref{note1} 
holds here too. 
For $\langle\theta^{\mu}\theta^{\rho}\rangle$ in (\ref{eq:321}) 
to be well-defined, either the gauge parameter $\alpha$ should be 
set 0 or the field $\theta_{\mu}$ should appear in company 
with derivatives in interaction Lagrangian. 
For $\langle\varphi\varphi\rangle$ in (\ref{eq:323}) to be 
well-defined, either the gauge parameter $\beta$ should be chosen 
as $\frac{1}{3}$ or the field $\varphi$ should  be accompanied 
by derivatives in interaction Lagrangian.
}
Izawa's procedure does work in this case.

\section{Massless limit}

It has been found that the theory constructed in the previous 
section has a smooth massless limit.
In this section, we investigate whether or not the limit is 
consistent with the ordinary massless tensor theory. 

As $m$ tends to 0, the Lagrangian $L_{{\rm T}, a=1}'$ 
(\ref{eq:307}) reduces to
\begin{eqnarray}
L & = & 
\frac{1}{2}h^{\mu\nu}\Lambda_{\mu\nu ,\rho\sigma}h^{\rho\sigma} 
+ 2\partial^{\mu}\varphi
\left(\partial^{\nu}h_{\mu\nu} - \partial_{\mu}h\right) 
+ b^{\mu}\left(
\partial^{\nu}h_{\mu\nu} - \frac{1}{2}\partial_{\mu}h 
+ \frac{\alpha}{2}b_{\mu}\right) 
\nonumber \\
& & \makebox[2cm]{}
- \frac{1}{2}\left(
\partial_{\mu}\theta_{\nu} - \partial_{\nu}\theta_{\mu}\right)^{2}
+ b\left(
\partial^{\mu}\theta_{\mu} + \frac{\beta}{2}b\right) ,
\label{eq:401}
\end{eqnarray}
where the trivial FP-ghost terms have been omitted.
The $\left( h_{\mu\nu}, \varphi, b_{\mu}\right)$-sector is 
completely separated from the 
$\left( \theta_{\mu}, b\right)$-sector. 
If it were not for the second term in the first line of the right 
hand side of (\ref{eq:401}), 
the $\left( h_{\mu\nu}, \varphi, b_{\mu}\right)$-sector coincides 
with the ordinary massless tensor theory. 
Two-point functions are for 
the $\left( h_{\mu\nu}, \varphi, b_{\mu}\right)$-sector 
\begin{eqnarray}
\langle h^{\mu\nu}h^{\rho\sigma}\rangle & = & 
\frac{1}{\Box}\left\{
\frac{1}{2}\left(
\eta^{\mu\rho}\eta^{\nu\sigma} 
+ \eta^{\mu\sigma}\eta^{\nu\rho} 
- \eta^{\mu\nu}\eta^{\rho\sigma}\right) \right. \nonumber \\
& & \mbox{}
- \frac{1}{2}(1-2\alpha )\frac{1}{\Box}\left(
\eta^{\mu\rho}\partial^{\nu}\partial^{\sigma}
+ \eta^{\mu\sigma}\partial^{\nu}\partial^{\rho}
+ \eta^{\nu\rho}\partial^{\mu}\partial^{\sigma}
+ \eta^{\nu\sigma}\partial^{\mu}\partial^{\rho}\right) \nonumber \\
& & \makebox[-1mm]{}\left.\mbox{}
+ \frac{2}{3}\left(
\frac{1}{2}\eta^{\mu\nu} 
+ \frac{\partial^{\mu}\partial^{\nu}}{\Box}\right)\left(
\frac{1}{2}\eta^{\rho\sigma} 
+ \frac{\partial^{\rho}\partial^{\sigma}}{\Box}\right)
\right\} \delta ,
\label{eq:402} \\ 
\langle h^{\mu\nu}b^{\rho}\rangle & = & 
\frac{1}{\Box}\left( 
\eta^{\mu\rho}\partial^{\nu} 
+ \eta^{\nu\rho}\partial^{\mu}\right)\delta , 
\label{eq:403} \\
\langle h^{\mu\nu}\varphi\rangle & = & 
\frac{1}{3}\frac{1}{\Box}\left(
\frac{1}{2}\eta^{\mu\nu} 
+ \frac{\partial^{\mu}\partial^{\nu}}{\Box}\right)\delta , 
\label{eq:404} \\
\langle b^{\mu}b^{\rho}\rangle & = & 0 , 
\label{eq:405} \\
\langle b^{\mu}\varphi\rangle & = & 0 , 
\label{eq:406} \\
\langle\varphi\varphi\rangle & = & 
\frac{1}{6}\frac{1}{\Box}\delta ,
\label{eq:407}
\end{eqnarray}
and for the $\left( \theta_{\mu}, b\right)$-sector 
\begin{eqnarray}
\langle\theta^{\mu}\theta^{\rho}\rangle & = & 
\frac{1}{2}\frac{1}{\Box}\left\{ \eta^{\mu\rho} 
- \left( 1 - 2\beta\right)
\frac{\partial^{\mu}\partial^{\rho}}{\Box}\right\}\delta ,
\label{eq:408} \\
\langle \theta^{\mu}b\rangle & = & 
\frac{1}{\Box}\partial^{\mu}\delta , 
\label{eq:409} \\
\langle bb\rangle & = & 0 . 
\label{eq:410} 
\end{eqnarray}
The $\left( h_{\mu\nu}, \varphi, b_{\mu}\right)$-sector does not 
reproduce the two-point functions for the ordinary massless 
tensor theory. 
If we didn't have the third line in the expression (\ref{eq:402}) 
for $\langle h^{\mu\nu}h^{\rho\sigma}\rangle$, 
and if $\langle h^{\mu\nu}\varphi\rangle$ were 0 in 
(\ref{eq:404}) , 
then the whole set of two-point functions agrees with that 
of the ordinary massless tensor.

In order to see how the ordinary graviton field is contained in 
this model, we now introduce the following nonlocal combination of 
the basic fields:
\begin{equation}
H_{\mu\nu} = h_{\mu\nu} 
- 2\left( \frac{1}{2}\eta_{\mu\nu} 
+ \frac{\partial_{\mu}\partial_{\nu}}{\Box}\right)\varphi .
\end{equation}
For this new field we have
\begin{eqnarray}
\langle H^{\mu\nu}H^{\rho\sigma}\rangle & = & 
\frac{1}{\Box}\left\{
\frac{1}{2}\left(
\eta^{\mu\rho}\eta^{\nu\sigma} 
+ \eta^{\mu\sigma}\eta^{\nu\rho} 
- \eta^{\mu\nu}\eta^{\rho\sigma}\right) \right. \nonumber \\
& & \left. \mbox{}
- \frac{1}{2}(1-2\alpha )\frac{1}{\Box}\left(
\eta^{\mu\rho}\partial^{\nu}\partial^{\sigma}
+ \eta^{\mu\sigma}\partial^{\nu}\partial^{\rho}
+ \eta^{\nu\rho}\partial^{\mu}\partial^{\sigma}
+ \eta^{\nu\sigma}\partial^{\mu}\partial^{\rho}\right) 
\right\}\delta , \nonumber \\
& &
\label{eq:412} \\ 
\langle H^{\mu\nu}b^{\rho}\rangle & = & 
\frac{1}{\Box}\left( 
\eta^{\mu\rho}\partial^{\nu} 
+ \eta^{\nu\rho}\partial^{\mu}\right)\delta , 
\label{eq:413} \\
\langle H^{\mu\nu}\varphi\rangle & = & 0  
\label{eq:414} \\
\end{eqnarray}
instead of (\ref{eq:402}), (\ref{eq:403}) and (\ref{eq:404}). 
Thus the two-point functions for the ordinary massless tensor 
field are in fact provided by the field $H_{\mu\nu}$.

\section{Summary and discussion}

It has turned out that: \\
(1) The original ASG model for a massive tensor field 
has second-order massless singularities, 
which can be regularized by applying Izawa's procedure 
based on BRS symmetry once; \\
(2) The original PT model develops fourth-order massless 
singularities, which can be regularized by applying Izawa's 
procedure twice.

The Batalin-Fradkin algorithm \cite{rf:3} is another powerful 
method for constructing gauge-invariant theories from 
non-gauge-invariant ones. 
The application of this procedure to a massive tensor field has 
been performed in Ref.~\citen{rf:4}, 
giving the same Lagrangians as obtained above, 
$L_{{\rm T}, a=\frac{1}{2}}$ (\ref{eq:208}) for 
the ASG model and 
$L_{{\rm T}, a=1}'$ (\ref{eq:307}) for the PT model.

We now have massive tensor theories equipped with BRS invariance 
as well as smooth massless limits.
However, we are still in the linearized world. 
To construct complete nonlinear theories is a major problem 
to solve.

\section*{Acknowledgements}
We would like to thank T. Kurimoto for discussion.

\appendix

\setcounter{section}{1}
\section*{Appendix \\
{\it --- On the Stueckelberg Formalism ---}}

It is known a massive vector field described by
\begin{equation}
L_{A} = \mbox{}-\frac{1}{4}F_{\mu\nu}F^{\mu\nu} 
- \frac{m^{2}}{2}A_{\mu}A^{\mu} 
\label{eq:A01}
\end{equation}
develops massless singularities in the limit $m=0$.
There are some methods to regularize such singularities. 
The most popular one is the Stueckelberg formalism.\cite{rf:5}\ In 
this formalism an additional scalar field $B$ is introduced 
and the Lagrangian and the physical state condition are given as
\begin{equation}
L_{\rm S} = \mbox{}
- \frac{1}{2}\partial_{\mu}A_{\nu}\partial^{\mu}A^{\nu} 
- \frac{m^{2}}{2}A_{\mu}A^{\mu} 
- \frac{1}{2}\partial_{\mu}B\partial^{\mu}B 
- \frac{m^{2}}{2}B^{2} , 
\label{eq:A02} 
\end{equation}
\begin{equation}
\left( \partial^{\mu}A_{\mu} + mB\right)^{(+)}|\;\rangle = 0 .
\label{eq:A03}
\end{equation}
On the other hand, as shown in I, 
Izawa's procedure based on BRS symmetry 
can also give a massless-regular theory, 
in which the Lagrangian is given by 
\footnote{
The same Lagrangian is also obtained by applying the 
Batalin-Fradkin 
algorithm.\cite{rf:6}
}
\begin{equation}
L_{\rm BRS} = \mbox{}-\frac{1}{4}F_{\mu\nu}F^{\mu\nu} 
- \frac{m^{2}}{2}\left(A_{\mu} - \frac{1}{m}\partial_{\mu}
\theta\right)^{2}
+ b\left(\partial^{\mu}A_{\mu} + \frac{\alpha}{2}b\right) 
+ i\bar{c}\Box c 
\label{eq:A04}
\end{equation}
with an auxiliary scalar field $\theta$, an NL field $b$, 
a pair of FP ghosts $(c,\bar{c})$, 
and a gauge parameter $\alpha$. 
This Appendix is devoted to show these two formulations are 
equivalent with each other. To do that we have to establish both 
the equivalence of path integrals and that of physical state 
conditions.

The path integral of the BRS-Izawa theory is
\begin{equation}
Z = \int {\cal D}A_{\mu}{\cal D}\theta{\cal D}b{\cal D}
c{\cal D}\bar{c}
\exp i\int d^{4}x\, L_{\rm BRS} .
\end{equation}
This is equivalent to 
\begin{equation}
Z = \int {\cal D}A_{\mu}{\cal D}\theta
\delta\left( \partial^{\mu}A_{\mu} - f\right){\rm Det}N
\exp i\int d^{4}x\, L_{\rm BRS}' ,
\label{eq:A06}
\end{equation}
where $L_{\rm BRS}'$ and $N$ are defined by
\begin{eqnarray}
L_{\rm BRS}' & = & \mbox{}
-\frac{1}{4}F_{\mu\nu}F^{\mu\nu} 
- \frac{m^{2}}{2}\left(A_{\mu} - \frac{1}{m}\partial_{\mu}
\theta\right)^{2}
\nonumber \\
& = & \mbox{}
- \frac{1}{2}\partial_{\mu}A_{\nu}\partial^{\mu}A^{\nu} 
- \frac{m^{2}}{2}A_{\mu}A^{\mu} 
- \frac{1}{2}\partial_{\mu}\theta\partial^{\mu}\theta 
- \frac{m^{2}}{2}\theta^{2} 
\nonumber \\
& & \makebox[15mm]{}
+ \frac{1}{2}\left(\partial^{\mu}A_{\mu} - m\theta\right)^{2} ,
\\
N & = & \Box\delta^{4}(x-x') ,
\end{eqnarray}
and $f$ is an arbitrary function of $x$.
The expression (\ref{eq:A06}) corresponds to the gauge-fixing 
condition
\begin{equation}
\partial^{\mu}A_{\mu} = f .
\end{equation}
We are allowed to take another gauge fixing condition
\begin{equation}
\partial^{\mu}A_{\mu} - m\theta = f' 
\end{equation}
with another arbitrary function $f'$. 
Because the Lagrangian $L_{\rm BRS}'$ is 
invariant under the gauge transformation 
\begin{equation}
\delta A_{\mu} = \partial_{\mu}\varepsilon , 
\makebox[1cm]{}
\delta\theta = m\varepsilon ,
\end{equation}
the factor 
\begin{equation}
\delta\left( \partial^{\mu}A_{\mu} - f\right){\rm Det}N
\end{equation}
in (\ref{eq:A06}) can be replaced by 
\begin{equation}
\delta\left( \partial^{\mu}A_{\mu} - m\theta - f'\right)
{\rm Det}N' 
\end{equation}
with 
\begin{equation}
N' = \left(\Box - m^{2}\right)\delta^{4}(x-x') .
\end{equation}
That is 
\begin{equation}
Z = \int {\cal D}A_{\mu}{\cal D}\theta
\delta\left( \partial^{\mu}A_{\mu} - m\theta - f'\right)
{\rm Det}N'
\exp i\int d^{4}x\, L_{\rm BRS}' .
\label{eq:A15}
\end{equation}
Taking into account the $f'$-independence for $Z$, 
we multiply (\ref{eq:A15}) by 
\begin{equation}
1 = \int{\cal D}f'\exp i\int d^{4}x\left(
-\frac{1}{2}f^{'2}\right) .
\end{equation}
Then we have 
\begin{eqnarray}
Z  & = & 
\int {\cal D}A_{\mu}{\cal D}\theta{\cal D}f'
\delta\left( \partial^{\mu}A_{\mu} - m\theta - f'\right)
{\rm Det}N'
\exp i\int d^{4}x\left(L_{\rm BRS}' - \frac{1}{2}f^{'2}\right) 
\nonumber \\
& = & 
\int {\cal D}A_{\mu}{\cal D}\theta
{\rm Det}N'
\exp i\int d^{4}x\, L_{\rm BRS}'' ,
\end{eqnarray}
where
\begin{equation}
L_{\rm BRS}'' = \mbox{}
- \frac{1}{2}\partial_{\mu}A_{\nu}\partial^{\mu}A^{\nu} 
- \frac{m^{2}}{2}A_{\mu}A^{\mu} 
- \frac{1}{2}\partial_{\mu}\theta\partial^{\mu}\theta 
- \frac{m^{2}}{2}\theta^{2} . 
\end{equation}
This Lagrangian $L_{\rm BRS}''$ shares the same form with 
the Stueckelberg Lagrangian $L_{\rm S}$ (\ref{eq:A02}). 
The equivalence of the path integrals has thus been confirmed.

Next come the physical state conditions. 
The expression (\ref{eq:A15}) is equivalent to
\begin{equation}
Z = \int {\cal D}A_{\mu}{\cal D}\theta
{\cal D}b{\cal D}c{\cal D}\bar{c}
\exp i\int d^{4}x\, L_{\rm BRS}''' ,
\label{eq:A19}
\end{equation}
where
\begin{eqnarray}
L_{\rm BRS}''' & = & \mbox{}
- \frac{1}{4}F_{\mu\nu}F^{\mu\nu} 
- \frac{m^{2}}{2}\left(A_{\mu} - \frac{1}{m}\partial_{\mu}
\theta\right)^{2}
\nonumber \\ 
& & \makebox[10mm]{}
+ b\left(\partial^{\mu}A_{\mu} - m\theta 
+ \frac{\alpha}{2}b\right) 
+ i\bar{c}\left(\Box - m^{2}\right) c .
\end{eqnarray}
Because this Lagrangian is invariant under the BRS 
transformation
\begin{equation}
\delta A_{\mu} = \partial_{\mu}c , 
\makebox[1cm]{}
\delta\theta = mc ,
\makebox[1cm]{}
\delta\bar{c} = ib ,
\end{equation}
a conserved BRS charge $Q_{\rm B}$ is defined. 
This is expressed as 
\begin{equation}
Q_{\rm B} = \int d^{3}x\, 
\left(b\,\partial_{0}c - \partial_{0}b\!\cdot\! c \right) .
\end{equation}
By the use of this charge, we can impose as usual the physical 
state condition
\begin{equation}
Q_{\rm B}|\;\rangle = 0 ,
\end{equation}
which is shown to be equivalent to the condition 
\begin{equation}
b^{(+)}(x)|\;\rangle = 0 
\makebox[7mm]{}{\rm for}\; \forall x .
\label{eq:A24}
\end{equation}
The field equation
\begin{equation}
\partial^{\mu}A_{\mu} - m\theta + \alpha b = 0 
\end{equation}
tells the condition (\ref{eq:A24}) reduces to 
\begin{equation}
\left(\partial^{\mu}A_{\mu}(x) 
- m\theta(x)\right)^{(+)}|\;\rangle = 0 
\makebox[7mm]{}{\rm for}\; \forall x .
\end{equation}
This agrees with the condition (\ref{eq:A03}). 
The equivalence of the physical state conditions has thus 
been confirmed too.

\end{document}